\documentclass[prb,twocolumn,showpacs,preprintnumbers]{revtex4}
\usepackage{graphicx}
\usepackage{dcolumn}

\begin{document}

\title{High-Pressure Phase Diagram in the Manganites: a Two-site Model Study}
\author{A. Sacchetti, P. Postorino}
\affiliation{Coherentia-INFM and Dipartimento di Fisica, Universit\`a ``La Sapienza'',
P.le A. Moro 4, I-00187 Roma, Italy.}
\author{M. Capone}
\affiliation{Enrico Fermi Center, Roma, Italy}

\begin{abstract}
The pressure dependence of the Curie temperature $T_C$ in manganites,
recently studied over a wide pressure range, is not quantitatively accounted
for by the quenching of Jahn-Teller distortions, and suggests the occurrence
of a new pressure-activated localizing processes. We present a theoretical
calculation of $T_C$ based on a two-site double-exchange model with
electron-phonon coupling interaction and direct superexchange between the $%
t_{2g}$ core spins. We calculate the pressure dependence of $T_C$ and
compare it with the experimental phase diagram. Our results describe the
experimental behavior quite well if a pressure-activated enhancement of the
antiferromagnetic superexchange interaction is assumed.
\end{abstract}


\date{\today}
\maketitle

\section{Introduction}

\label{Intro}

Rare-earth manganites (A$_{1-x}$A$_{x}^{\prime }$MnO$_{3}$, where A is a
trivalent rare earth and A$^{\prime }$ a divalent alkali earth) have been
the object of renewed interest owing to the discovery of the Colossal
Magneto-Resistance (CMR) \cite{Jin} exhibited by several of these compounds
over the $0.2\leq x\leq 0.5$ doping range.
The physics underlining the properties of CMR-manganites is very rich and
not yet completely understood,\cite{Cheong} although CMR is commonly
described in the framework of the Double-Exchange (DE) model.\cite{Zener}
DE qualitatively accounts for the closely related phenomena of CMR and
temperature-driven transition from the paramagnetic insulating phase (high
temperature) to the ferromagnetic metallic one (low temperature).
Nevertheless, the remarkable quantitative disagreement with the experimental
data indicates that additional and different physical effects must be
introduced.\cite{Millis1} The key role of the Jahn-Teller (JT) effect, i.e.,
a spontaneous distortion of the MnO$_{6}$ octahedron, which reduces the
energy of the system by lifting the degeneracy of the $e_{g}$ levels, was
pointed out by Millis et al. \cite{Millis1} and it is nowadays widely
accepted. This effect is naturally associated with a sizeable
electron-phonon (el-ph) coupling, which reduces the electron mobility, leading
to the formation of small polarons and competing with the delocalizing DE
mechanism. A complete description of manganite properties at microscopic
level is further complicated by the presence of an antiferromagnetic (AF)
Super-Exchange (SE) interaction between the $S=3/2$ spins of the localized
$t_{2g}$ electrons, which is
relevant in the electron-doped regime ($x>0.5$) and could play a role also
in the CMR region.

High pressure is a powerful tool to investigate the role of the JT effect in
manganites. The application of an external pressure produces a
symmetrization of the MnO$_{6}$ octahedra, thus reducing the lattice
distortions and enhancing carriers mobility. Therefore, applied pressure
causes an increase of the metal-insulator transition temperature $T_{IM}$,
which in these systems is almost coincident with the Curie temperature 
$T_{C} $. An almost linear increase of the transition temperature 
$T_{C}$ with pressure $P$ has been observed in several manganites over a
moderate pressure range (0-2 GPa).\cite{Hwang,Laukhin}
In a recent paper the effect of pressure on the metal-insulator transition
temperature in La$_{0.75}$Ca$_{0.25}$MnO$_{3}$ ($x=$ 0.25) has been studied
up to 11 GPa by means of optical measurements.\cite{PRL} The results
clearly show that the linear dependence of $T_{C}(P)$ is limited only to the
low pressure regime ($P< 3$ GPa), whereas in the high-pressure regime ($P>
6$ GPa) the $T_{C}(P)$ curve bends down and it seems to approach an
asymptotic value. This behavior, consistently with previous results from
high-pressure spectroscopic measurements,\cite{PRLold,PRBold} has been
ascribed to the onset of a localizing, pressure-activated, mechanism
competing with the natural pressure-induced charge delocalization.

The present paper aims at explaining the experimental $T_{C}(P)$ behavior
using a simple theoretical approach which enables a deep understanding of
the microscopic mechanisms driving the observed pressure behavior. 
We propose a model which contains, besides the \textit{standard} DE and el-ph
coupling, 
also AF-SE coupling between the $t_{2g}$ core spins, which may provide the 
new localizing mechanism at high-$P$ invoked to account
for the experimental behavior.\cite{PRL,PRLold,PRBold} 
It has been proposed that the competition between AF-SE and DE accounts for
some of the magnetic properties of the manganites.\cite{feinberg} The possible
role of the AF-SE term in providing the high-pressure extra localizing channel
is, indeed, quite reasonable, since the superexchange between $t_{2g}$
orbitals may be thought as proportional to the square of some hopping
integral $t_{t}$ between the $t_{2g}$ orbitals of neighboring manganese
ions. The hopping integral, arising from the overlap of the atomic
wavefunctions is expected to increase with increasing pressure. 

An extremely simple, yet efficient, model to compute $T_{C}(P)$ is the two
Mn-site cluster, which represents the minimal model including DE, el-ph
coupling, and
SE interactions. The simplicity of this model enables to carry out exact
calculations at finite temperature, which are fundamental to describe the
experimental $P-T$ phase diagram. The two-site model (TSM) here presented has
already been applied in Ref. \onlinecite{Capone}, where it has been shown to
successfully catch the relevant physics of manganites at $T=0$ owing to the
extremely short-range nature of the interactions occurring in these systems. 
Here we extend
the study to finite temperature and discuss the dependence on pressure of
the various parameters in the Hamiltonian. In this way we treat, at the same
level, the quenching of the JT interactions and the role of AF-SE coupling
in determining the pressure dependence of the metal-insulator transition. We
show that the experimental results are well described if the AF-SE
interaction is assumed to be renormalized proportionally to the square of
the hopping matrix element.

The paper is organized as follows: in Sec. \ref{Modcal} we introduce the model and 
the details of the calculation; In Sec. \ref{TcP} the pressure dependence of $T_{C}$
is reported and discussed, while Sec. \ref{conc} is dedicated to the conclusions.

\section{Model and Calculations}

\label{Modcal}

The TSM has been extensively employed in the study of polaronic systems,
since it is the simplest model able to describe the crossover from
a metal to an almost localized small polaron, which is reflected in both
ground state  \cite{Depasq,Acquar} and the spectral properties.\cite{Rann,Demel} 
The DE-TSM including el-ph coupling has been already studied 
at $T=0$ \cite{Capone} and its relevance to the case of CMR
manganites has been discussed.\cite{Capone,Satp1,Mitra} In particular, the
correct ground state has been obtained in several coupling regimes. The
qualitative difference between quantum ($S$=3/2) and classical ($S=\infty )$
spin cases was also pointed out, showing the importance of quantum
fluctuations of the core-spins in a proper study of manganites.\cite{Capone} 
In the present paper, we extend the model of Ref. \onlinecite{Capone} to finite
temperature and focus our attention on the temperature-driven crossover from
paramagnetic to ferromagnetic ordering occurring in manganites. In such a
minimal cluster no phase transition with long-range order can occur but,
owing to the very short-range nature of the interactions in manganites,
predictions on possible instabilities taking place in thermodynamic limit
can be inferred.

The hamiltonian of our TSM reads:

\begin{eqnarray}  \label{hamiltonian}
H=-t \sum_{\sigma} {\left( c_{1,\sigma}^{\dagger} c_{2,\sigma} +
c_{2,\sigma}^{\dagger} c_{1,\sigma} \right)} - J_H \sum_{i=1,2} 
{\vec{\sigma}_i \cdot \vec{S}_i}  \nonumber \\
+ J_1 \vec{S}_1 \cdot \vec{S}_2 - g \left( n_1 - n_2 \right) (a+a^\dagger ) + 
\omega_0
a^{\dagger} a.
\end{eqnarray}

\noindent In the electronic part (first three terms), $t$ is the hopping
integral between the $e_{g}$ levels, $c_{i,\sigma }^{\dagger }$ 
($c_{i,\sigma })$ is the creation (annihilation) operator for an electron of
spin $\sigma $ on site $i$, $J_{H}$ is the Hund's rule coupling, $\vec{\sigma}
_{i}=c_{i\alpha }^{\dagger }\vec{\sigma}_{\alpha \beta }c_{i\beta }$ is the
spin operator on site $i$ ($\vec{\sigma}_{\alpha \beta }$ are the Pauli
matrices), $\vec{S}_{i}$ is the local spin ($\left| {\vec{S}_{i}}\right|
=3/2)$ due to localized $t_{2g}$ core electrons on site $i$, and $J_{1}$ is 
the AF-SE coupling.
The phonon contribution (last two terms) contains an Holstein coupling and an
harmonic term, where $g$ is the electron-phonon coupling, 
$n_{i}=\Sigma _{\sigma }c_{i,\sigma }^{\dagger }c_{i,\sigma }$ is the
electron number operator at site $i$, and 
$a^{\dagger }$ ($a$) creates (annihilates) an Einstein phonon of frequency 
$\omega _{0}$, coupled to the density difference between the two sites. The
above form of the el-ph coupling is equivalent to local phonons coupled to the 
on-site electron density $n_{i}$, after eliminating the symmetric phonon mode 
which couples to the total density.\cite{Capone} For the sake of simplicity, 
we replace the JT coupling with the standard Holstein el-ph coupling.
It has been shown that, as far as the evaluation of $T_{C}$ is concerned, 
the Holstein coupling in a single orbital model gives results
extremely close to a two-orbital model with JT interactions.\cite{Roder}
The independent parameters of this model are $J_{H}/t$, $J_{1}/t$, $\omega
_{0}/t$, and the dimensionless el-ph coupling $\lambda =2g^{2}/\omega _{0}t$. 
As already mentioned, the extreme simplicity of the model does not require 
further approximations and allows for an exact solution for arbitrary values 
of the parameters. As an example, the quantum nature of phonons and core spin 
can be fully taken into account.

In the present calculation we will consider the case of a single electron on
the two sites, for which the Hamiltonian (\ref{hamiltonian}) has a very
small ($64 \times 64$) electronic Hilbert-space (including the quantum $S=3/2$ 
spins of the $t_{2g}$ electrons), while the infinite phonon
Hilbert space necessarily requires a truncation up to some maximum phonon
number $n_{ph}$. Except from the extreme adiabatic case ($\omega _{0}/t\ll 1$) 
and/or extreme multiphononic regime ($\lambda t/\omega_0 \gg 1$), where a
really large number of phonons can be excited, convergence of the relevant
states can be, in general, achieved for relatively small ($<$100) $n_{ph}$.
In the non-adiabatic and intermediate coupling regime a complete
diagonalization of (\ref{hamiltonian}) can be carried out numerically. In
the case of manganites, where we can estimate
$\omega _{0}/t\approx $0.5 and $\lambda \approx $ 1 (see below),
the convergence for the low energy states is achieved already at $n_{ph}$=10.

The complete diagonalization of $H$, i.e., the determination of all
eigenvalues $E_{n}$ and eigenvectors $\left| {\psi _{n}}\right\rangle $,
enables the calculation of the thermal average of the 
nearest-neighbor spin correlation operator 
$\mu =(\vec{S}_{1}\cdot \vec{S}_{2})/S^{2}$ as 

\begin{equation}
\mu (T)=\frac{\sum\limits_{n}{\left\langle {\psi _{n}}\right| \mu \left| 
{\psi _{n}}\right\rangle \exp \left( {-E_{n}/T}\right) }}{\sum\limits_{n}
{\exp \left( {-E_{n}/T}\right) }}.  \label{mut}
\end{equation}

\noindent This quantity measures the short-range magnetic correlations and
it is positive for ferromagnetic phases, negative for antiferromagnetic, and
vanishes in the high-temperature paramagnetic phase. As we will show in the
next section, a simple analysis of the temperature dependence of the
spin-correlation $\mu (T)$, allows us to give a reliable estimate of $T_{C}$.
The choice of this ``short-range'' estimator of $T_C$ is, in our opinion,
preferable to thermodynamic estimates in our TSM.
The reliability of the TSM in the case
of manganites can be assessed also through the analogy with a
completely different theoretical approach, the Dynamical Mean Field Theory
(DMFT),\cite{Georges} which has been applied to a lattice model for the
manganites by Millis and coworkers.\cite{Millis2} The DMFT is a powerful
nonperturbative approach which freezes spatial fluctuations, but fully
retains the local (single-site) quantum dynamics, and becomes exact in the 
limit of infinite coordination.\cite{Georges} The DMFT maps the original 
lattice model in the thermodynamic limit onto a self-consistent impurity model 
which interacts with a quantum bath. As already discussed in 
Ref. \onlinecite{Capone}, we emphasize the conceptual analogy of these two 
approaches. In the TS model, the quantum nature of the problem is also 
completely retained, and each of the two sites ``feels'' the presence of the 
other site similarly to the way the impurity site feels the bath within DMFT. 
We stress that, within the DMFT approach, the system is in the thermodynamic 
limit. Both the methods, despite their differences, are thus expected to well 
describe the physics of short-range correlations in manganites. The
choice of the TSM allows us to easily include the nearest-neighbors
SE-AF interaction which is instead unaccessible by the single-site
DMFT and requires an Extended-DMFT
study, where the dynamical spin correlation function is also
self-consistently evaluated,\cite{Georges} or a cluster-DMFT approach,
which retains short-range dynamical correlations, but substantially
increases the computational effort.\cite{cdmft}

\section{Pressure dependence of $T_C$}

\label{TcP}

\begin{figure}[tb!]
\begin{center}
\includegraphics[width=8cm]{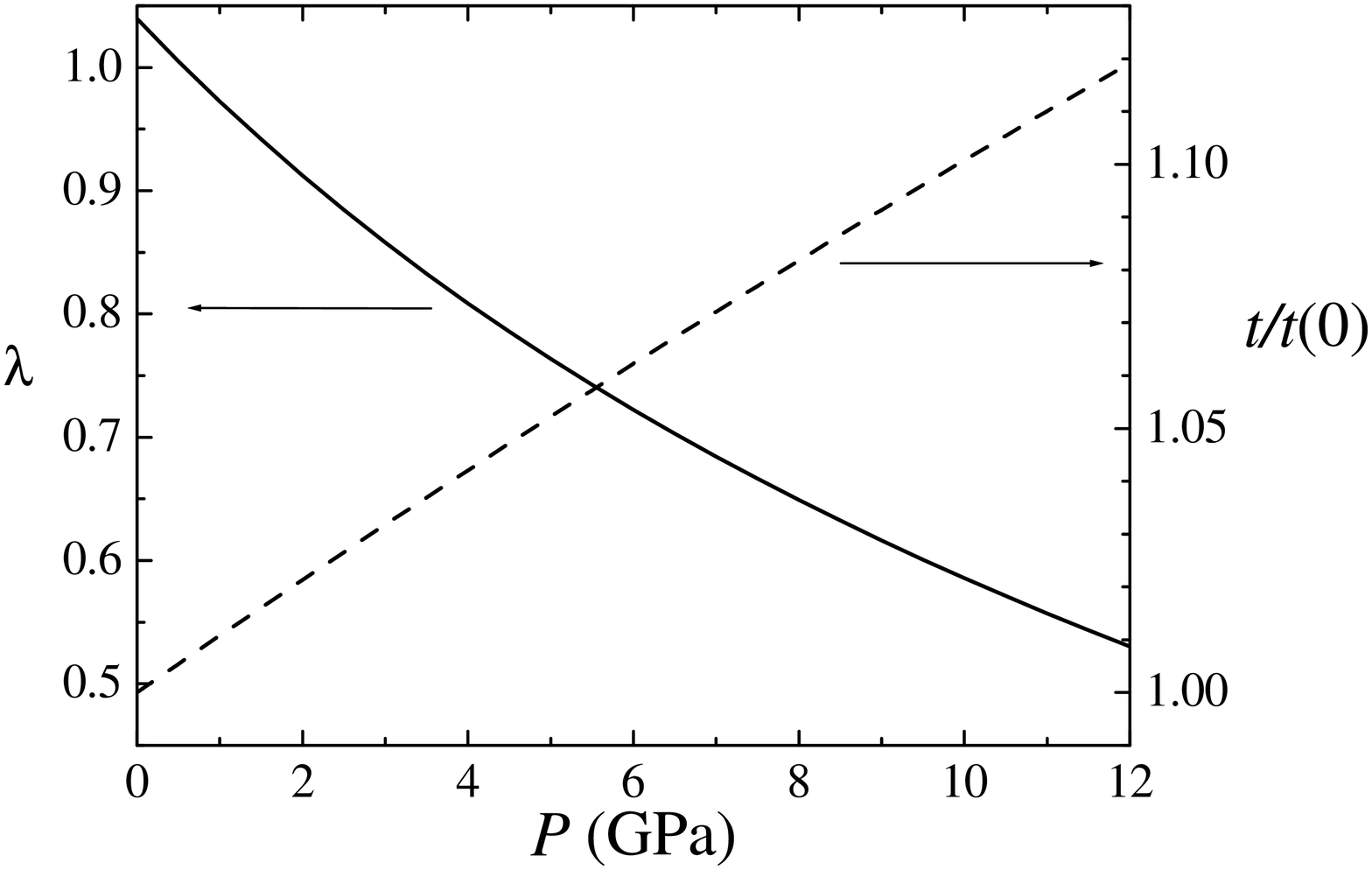}
\end{center}
\caption{Pressure dependence of the hopping integral (dashed line) and of the dimensionless electron-phonon coupling (solid line, see text).}
\label{ltvsp}
\end{figure}

In order to carry out the comparison between the experimental results
reported in Ref. \onlinecite{PRL} and the $T_{C}(P)$ values which can be
determined using the TSM, we need to know the pressure dependence of
the input model parameters $t(P)$, $J_{H}(P)$, $\omega _{0}(P)$, $\lambda
(P) $, and $J_{1}(P)$ for La$_{0.75}$Ca$_{0.25}$MnO$_{3}$ at least over the
0-11 GPa pressure range.
In Ref. \onlinecite{Liu} the hopping integral $t$ and the Hund's coupling 
$J_{H}$ for La$_{0.67}$Ca$_{0.33}$MnO$_{3}$ ($x=$ 0.33) have been derived from 
first principles calculations as a function of $a/a_{0}$, where $a$ is the 
cubic lattice parameter and $a_{0}$ its ambient pressure value. 
It is reasonable to assume the same dependence of $t$ upon $a$ in the 
$x=0.33$ and $x=0.25$ materials up to an overall factor, namely 
$t_{x=0.25}(a/a_{0})=Ct_{x=0.33}(a/a_{0})$. The proportionality 
factor $C$ can be evaluated
exploiting the proportionality between $t$ and $T_{C}$ at $P=0$ [Ref. \onlinecite{Cheong}],
i.e. through the relation $t_{x=0.25}/t_{x=0.33}=T_{C}(x=0.25)/T_{C}(x=
0.33)$. The pressure dependence of the Hund's coupling $J_{H}$ 
can be safely assumed to be identical for $x=0.25$ and $x=0.33$ compounds,
since $J_{H}$ is much larger than the other energy scales, and the results of 
the calculation are therefore only very weakly dependent on its variations. 
Finally, using the lattice
parameter $a(P)$ recently measured over the 0-15 GPa pressure range for 
La$_{0.75}$Ca$_{0.25}$MnO$_{3}$,\cite{Meneg} the $a/a_{0}$ dependence can be
simply converted into a pressure dependence. 
The increase of the hopping integral $t(P)$ as a function of pressure 
is plotted in Fig. \ref{ltvsp} in the relevant pressure range
(about 12 \% at 12 GPa).

The pressure dependence of the JT phonon frequency $\omega_{0}(P)$,
directly obtained from Raman measurements carried out over the 0-14 GPa
range,\cite{PRLold} is also useful to get an estimate for $\lambda (P)$. 
For a JT mode, it is reasonable to assume that the electron-phonon
coupling $g$ is given by $g=\sqrt{2M\omega_0} dt(a)/da$, where $M$ is the 
ionic mass. Using the above set of
pressure dependent parameters, the estimate for the pressure dependence of
the el-ph coupling $\lambda = 2g^{2}/\omega_{0}t$ can be obtained. The 
resulting $\lambda (P)$ is plotted in Fig \ref{ltvsp}. 
The zero-pressure value thus obtained, $\lambda(0)=1.04$, can be checked 
against an independent estimate of $g$. The typical Jahn-Teller distortion
$x_0 = \langle x\rangle = 1/\sqrt{2M\omega_0} \langle(a+a^{\dagger})\rangle$ 
can be estimated, in the polaronic regime, as $x_0 = \sqrt{2/M\omega_0} 
g/\omega_0$. Using the value $x_{0} =0.09$ \AA \ for 
La$_{0.75}$Ca$_{0.25}$MnO$_{3}$ at 
$P=0$ [Ref. \onlinecite{Lanzara}], one has $\lambda (0)=0.94$, which compares well
with the previous estimate. The remarkable
pressure-induced reduction of $\lambda (P)$ (about 50{\% }at 12 GPa), is
consistent with the observed enhancement of the metallic character of the
system and with the increase of the transition temperature $T_{C}$.\cite
{PRL,PRLold}
The last ingredient is the AF-SE coupling $J_{1}$ with its pressure
dependence. To our knowledge, no direct estimate of this parameter, either
experimental or theoretical, is available. An estimate of the zero-pressure 
value of $J_{1}(P)$ can be obtained by
simply imposing the calculated $T_{C}(0)$ be equal to the known experimental
value of 220 K for La$_{0.75}$Ca$_{0.25}$MnO$_{3}$ at zero pressure.\cite
{Cheong} 

\begin{figure}[tb!]
\begin{center}
\includegraphics[width=8cm]{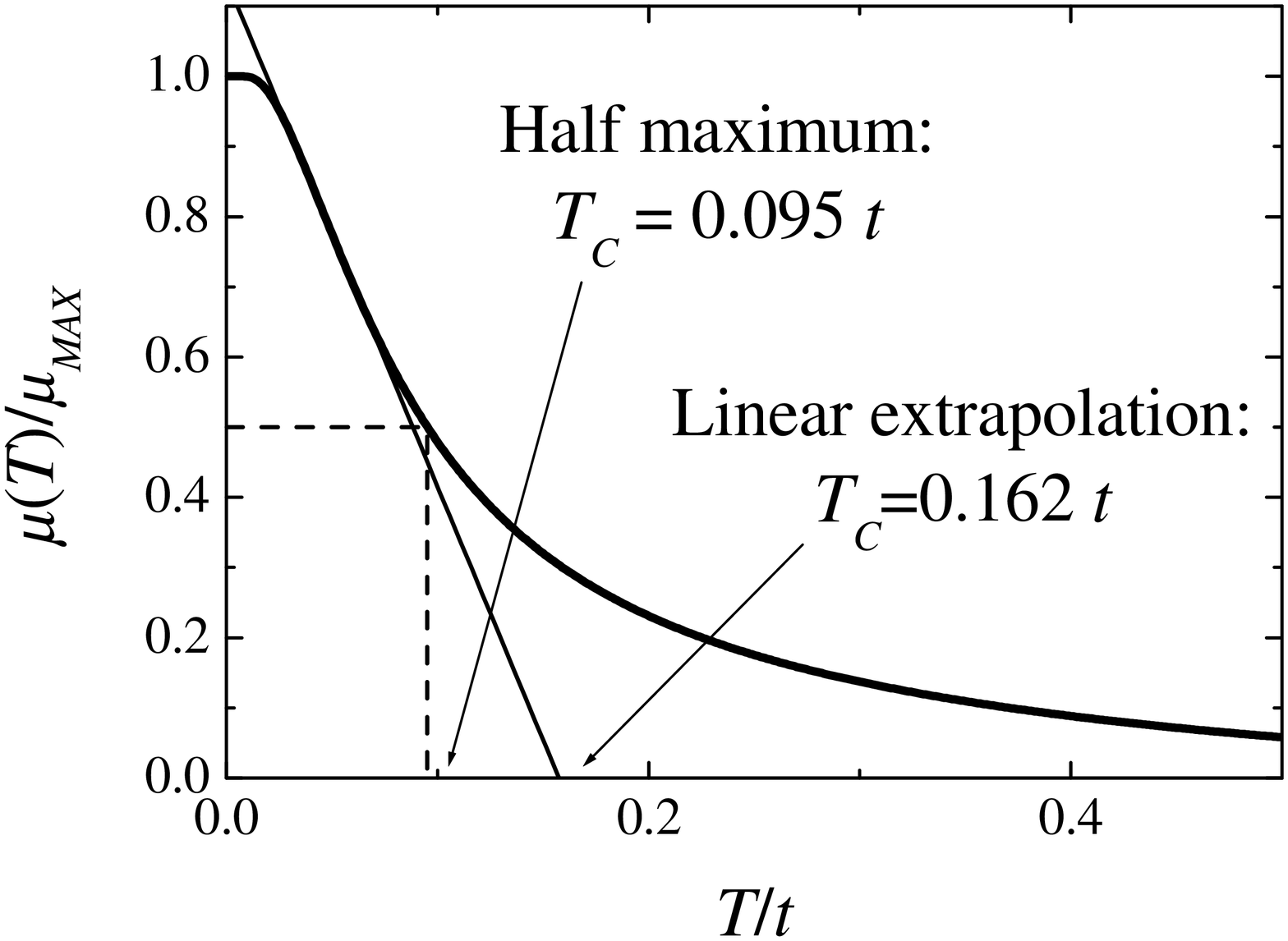}
\end{center}
\caption{Temperature dependence of the spin-correlation (thick solid line), calculated for $\lambda=$1.04, $J_H/t=$12.8, $\omega_0/t=$0.39, and $J_1/t=$0.047. Light solid and dashed lines represent the half-maximum and linear extrapolation methods for determining $T_C$, respectively (see text).}
\label{SSvsT}
\end{figure}

In Fig. \ref{SSvsT} we show $\mu (T)$ calculated in the TSM
using the known zero pressure values of the parameters ($t(0)\approx 0.20$
eV, $\omega _{0}(0)/t(0) = 0.39$, $J_{H}(0)/t(0) = 12.8$, $\lambda
(0)=1.04)$ and a value of $J_{1}(0)=0.047t$ which is consistent
with the experimental value of $T_{C}$. The crossover from a
ferromagnetic state at low temperature, with a large positive value of $\mu $, 
to a paramagnetic state at high temperature with $\mu \rightarrow 0$ is evident. 
The absence of an abrupt variation of $\mu (T)$ is expected since the real
phase transition occurring in the thermodynamic limit cannot take place in a
finite system and it is replaced by a smooth crossover. In the TSM the
crossover is quite broad and the estimate of the critical temperature is not
completely straightforward and implies some degree of arbitrariness.
For this reason, we propose, and compare, two different methods for
evaluating $T_{C}$. The first one, the half-maximum method, simply defines
the $T_{C}$ as the temperature for which $\mu (T_{C})=[\mu (0)-\mu (\infty)]/2
=\mu (0)/2$, i.e., for which the spin-correlation becomes one half of its
zero-temperature value. The second method is based on a linear extrapolation
around the inflection point of the $\mu (T)$ curve. These two methods are
schematically represented in Fig. \ref{SSvsT} from which it is apparent
that the latter method provides a higher value for $T_{C}$. Interestingly,
the two estimates of $T_{C}$ are actually
proportional over a wide range of parameter values. Therefore, the choice of
one method basically affects only the absolute value of $T_{C}$ and not the
internal comparison between results obtained with a given estimator. In the
following, we use the half-maximum estimate to determine the effect of
pressure on $T_{C}$.
It is important to note that the precise $J_{1}(0)$ value is influenced by
the method chosen for determining $T_{C}$. For example, the determination of 
$T_{C}$ with the half-maximum method leads to $J_{1}(0)=0.047t$(0) whereas
the linear extrapolation provides $J_{1}(0)\approx 0.060t(0)$. 
The slight difference between the zero pressure values is not important since 
we are mostly interested in the pressure dependence of the parameters rather
than in their absolute values.

\begin{figure}[tb!]
\begin{center}
\includegraphics[width=8cm]{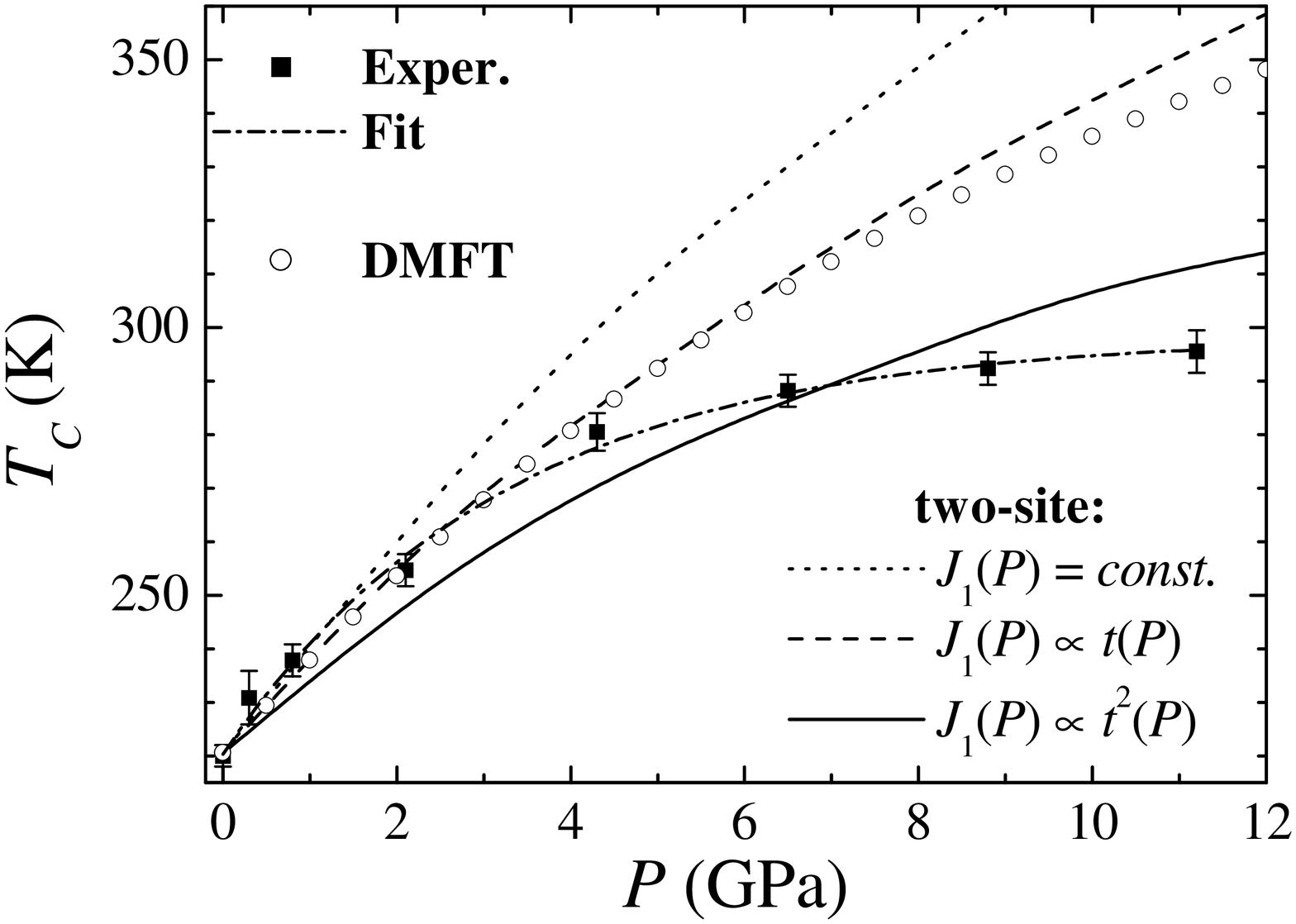}
\end{center}
\caption{$T_C(P)$ calculated with the two-site model for $J_1=const.$ (solid line), $J_1 \propto t$ (dashed line), 
and $J_1 \propto t^2$ (dotted line), compared with experimental values (squares). Empty circles represent the DMFT estimate \cite{Millis2} (see text). Dashed-dotted line is an exponential fit of experimental data \cite{PRL}.}
\label{tcvsp}
\end{figure}

As mentioned before, since the pressure dependence of $J_{1}$ is unknown, we
have carried out calculations of $T_{C}(P)$ for three different
pressure-dependences of $J_{1}(P)$ namely: i) $J_{1}(P)=const.$; ii) 
$J_{1}(P)\propto t(P)$; iii) $J_{1}(P)\propto t^{2}(P)$. Bearing in mind 
that  $t(P)$ is an increasing function of the pressure
(see Fig. \ref{ltvsp}), the three choices correspond
to reduce, keep constant, and increase the ratio $J_{1}/t$ upon increasing
pressure. The results of the TSM calculations together with the experimental
data from Ref. \onlinecite{PRL} are shown in Fig. \ref{tcvsp}. 
Before discussing the accuracy of our results with respect to the experiment,
we compare with previous DMFT results of Ref. \onlinecite{Millis2}, in which
a DE term with classical core spins, a JT-type el-ph interaction with classical
phonons and no AF-SE term have been used.
By combining the $T_{C}(\lambda )/t$ values reported for $n=0.75$ (i.e., 
$x=0.25$)\cite{Millis2} with the present estimate for $\lambda (P)$ and $t(P)$,
the $T_{C}(P)$ are readily obtained. The $T_{C}(P)$ values, also shown
in Fig. \ref{tcvsp}, have been divided by a factor 1.1 in order to force the
agreement with experimental data at zero pressure. Such a small difference
for the $T_{C}(0)$ value confirms the reliability of the present model
parameter estimates.
Comparing the theoretical results shown in Fig. \ref{tcvsp}, it is
interesting to note that DMFT and the TSM with $J_{1}\propto t(P)$
provide extremely close results. Beside the previously discussed
relationship between the two approaches, this agreement can be explained as
follows: The pressure dependence in the DMFT approach is basically
determined by $\lambda $, while in our TSM $\lambda $, $\omega _{0}$,
and $J_{1}$ depend on $P$. Since $\omega _{0}/t$ is only weakly dependent on
pressure, the choice $J_{1}\propto t$ leads to a TSM in which the
pressure dependence is substantially determined only by $\lambda $ as in the
DMFT estimate. This argument explains the observed equivalence between the
two approaches and, above all, it points out how the pressure dependence
of the el-ph interaction alone does not provide a satisfactory description of 
the observed experimental high-pressure behavior.

The comparison between our theoretical estimates and experimental data is
reported in  Fig. \ref{tcvsp}. The pressure dependence of the AF interaction 
remarkably affects the calculated $P-T$ phase diagram. For small pressures, 
the cases i) [$J_1=const$] 
and ii) [$J_1 \propto t$] nicely follow the experimental $T_C$,
while case iii) [$J_1 \propto t^2$] gives a lower critical temperature.
By increasing the pressure, in the two first cases $T_C$ increases too
rapidly with pressure, while the third estimate closely follows the 
experimental data up to 11 GPa. In particular,  $J_{1}=const$ produces an 
almost linear behavior of $T_{C}(P)$ over the whole pressure range.
The ability of the $J_1 \propto t^2$ data to reproduce the actual
experimental behavior is really encouraging, since the $t^{2}$ dependence may
be seen as the most ``physical'' approximation for $J_{1}$. The basic
hypothesis is that the pressure dependence of the overlap between the 
$t_{2g} $ orbitals is similar to the dependence of the same overlap for the 
$e_{g}$ orbitals, i.e., that for each pressure $t_{t}(P) \propto t(P)$. Under
this assumption, the superexchange between core spins $J_{1}\propto
t_{t}^{2}/U$ (where $U$ is an effective repulsion energy depending on $J_{H}$, 
the Mn-O charge transfer energy, and the on-site Coulomb repulsion)
becomes in turn proportional to $t^{2}/U$, $t$ being the hopping integral
between the $e_{g}$ orbitals.
These findings clearly show that the effects of pressure are not only
limited to a reduction of the el-ph coupling but a remarkable pressure 
dependence of
the AF interaction has to be introduced to account for the experimental
behavior. The same information can be obtained by extracting an 
``experimental'' dependence of the antiferromagnetic term on pressure $J_1(P)$.
This simply amounts to determine, for each pressure value, the value of
$J_1$ which provides the experimental $T_C$, given that all the other
parameters are fixed by the above mentioned conditions.

\begin{figure}[tb!]
\begin{center}
\includegraphics[width=8cm]{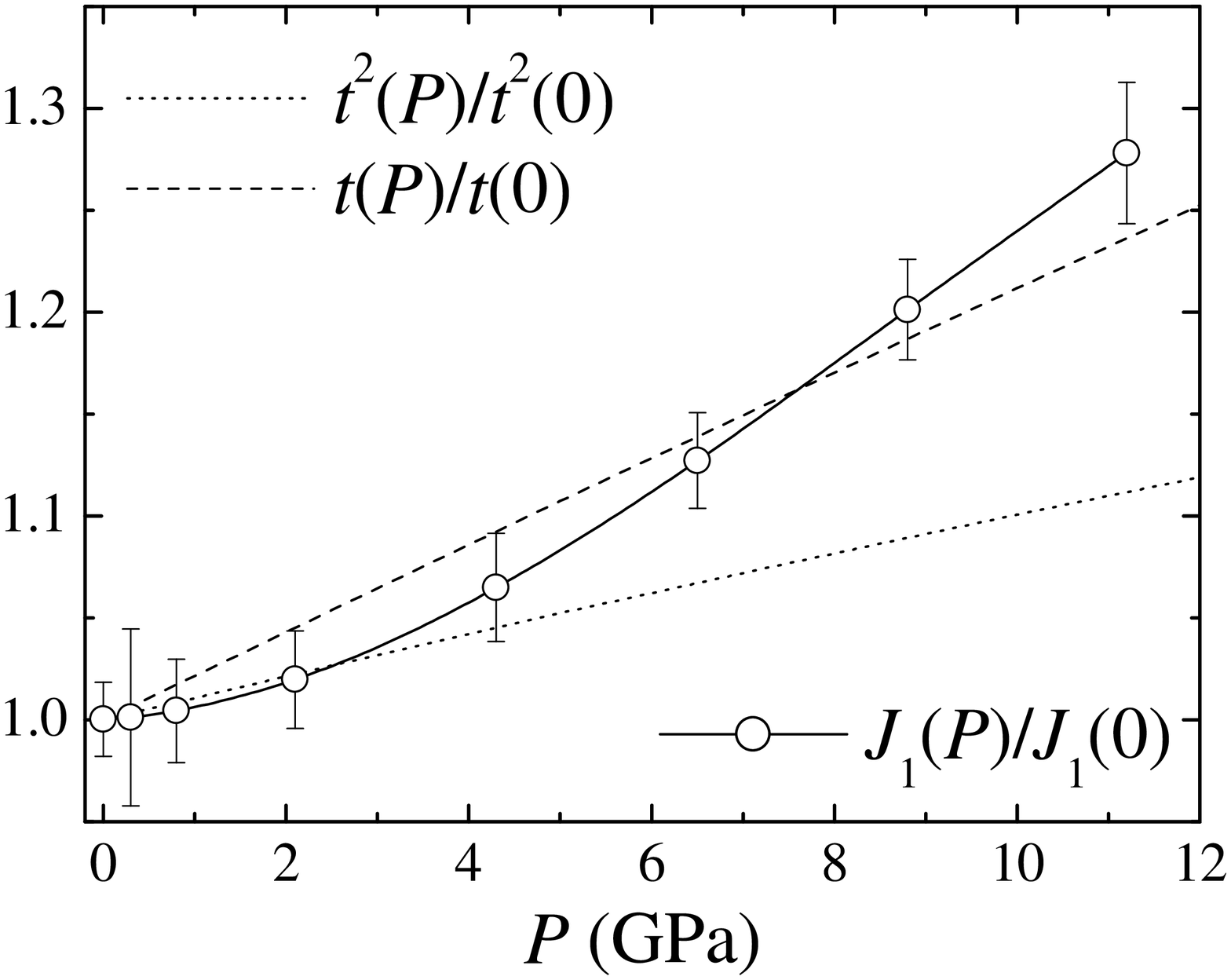}
\end{center}
\caption{Empty circles: pressure dependence of $J_1$ extracted from the experimental data of Ref. \onlinecite{PRL} by means of the two-site model (see text). Solid line is a guide to eyes. For sake of comparison $t^2(P)$ (dashed line) and $t(P)$ (dotted line) are also shown. All curves are normalized to their zero-pressure value.}
\label{j1optimal}
\end{figure}

In Fig. \ref{j1optimal}, we show $J_{1}(P)$, normalized to the zero-pressure 
value together with $t^{2}(P)/t^{2}(0)$. The two quantities compare quite well, 
strengthening the $J_{1}(P)\propto t^{2}(P)$ ansatz. For sake of comparison the 
quantity $t(P)/t(0)$ corresponding to the case ii)\ is also shown in
Fig. \ref{j1optimal}.
The above results strengthen the hypothesis that the saturation of 
$T_{C}(P)$ observed above 6-7 GPa \cite{PRL} can be ascribed to the onset of
a regime in which the AF interaction is no more negligible and competes with
the DE. This hypothesis is supported by several recent high-pressure
experiments. The occurrence of AF interactions in a DE system leads to charge
localization, which in turn may favor the coherence of the JT distortions.
This picture is consistent with high-pressure X-ray diffraction experiments,
in which the onset of a coherent JT distortion has been observed at about 7
GPa in La$_{0.75}$Ca$_{0.25}$MnO$_{3}$ and other compounds of the 
La$_{1-x}$Ca$_{x}$MnO$_{3}$ series.\cite{Meneg,Meneg2} It is also worth to 
notice that the onset of the AF phase in LaMnO$_{3}$ is accompanied by a remarkable
softening of the $B_{2g}$-JT phonon.\cite{Podo,Gran} Even if in our case no
real AF ordering occurs, the enhancement of the AF interaction suggested by
our analysis is consistent with the observed saturation of the same phonon
in La$_{0.75}$Ca$_{0.25}$MnO$_{3}$ at about 7 GPa.\cite{PRLold}

The $J_{1}(P)\propto t^{2}(P)$ dependence suggests that, upon increasing the 
pressure, the system evolves from a regime dominated by DE, where the bond
compression leads to an increase of $T_{C}$, to an intermediate regime,
where $T_{C}$ is almost constant and independent on pressure, owing to the
competition between SE and DE, and eventually to a very high pressure regime
dominated by the SE contribution where $T_{C}$ starts decreasing. This
picture is absolutely consistent with the pressure dependence of $T_{C}(P)$
recently observed in several manganites where the three regimes are apparent, 
\cite{Cui1,Cui2,Cui3} as well as with the data of Ref. \onlinecite{PRL} 
for La$_{0.75}$Ca$_{0.25}$MnO$_{3}$, where the system only reaches the 
intermediate regime for the maximum pressure. The extent of the
pressure range over which $T_{C}$ is almost pressure independent is
different for the samples studied in Refs. \onlinecite{Cui1,Cui2,Cui3}, 
although the onset of saturation takes place around 4 GPa for all these 
samples. This
pressure value is lower than that observed in La$_{0.75}$Ca$_{0.25}$MnO$_{3}$
\cite{PRL} and the difference can be ascribed to the presence of strong
coupling and/or high cation disorder which have been shown to have strong
effects on the pressure dependence of $T_{C}$.\cite{PRL}

\section{Conclusion}

\label{conc}

We studied the evolution of the transition temperature of 
La$_{0.75}$Ca$_{0.25}$MnO$_{3}$ with pressure by exactly solving a two Mn-site
model which contains all the relevant microscopic interactions acting
in the manganites. 
The theoretical results have been compared with the phase diagram recently
measured over the 0-11 GPa pressure range.\cite{PRL} 
As input parameters for
the model we used pressure dependent data available in literature, except
for the direct AF-SE between the $t_{2g}$ spins, for which, in absence
of experimental estimates, we compared three different pressure dependences.
We have shown that the pressure dependence law for $J_{1}$ is indeed crucial to
obtain a good agreement with the experiment. 
Neglecting the effect of pressure on the AF-SE term leads in fact to a sizeable
overestimate of the Curie temperature for $P > 6$ GPa. 
On the other hand, the theoretical results follow the experimental
data over the whole pressure range for $J_1(P) \propto t^2(P)$,
while the calculation largely overestimates the critical temperature
at large pressures if a weaker pressure dependence of $J_1$ is assumed.

The present results give a clear indication that the observed high pressure
behavior of $T_{C}$ can be ascribed to a pressure-induced enhancement of the
AF-SE which enters in competition with the DE above 6-7 GPa.
The idea of a
pressure activated AF interaction is, at least, compatible with the previous
experimental evidences of the onset of cooperative JT effect 
\cite{Meneg,Meneg2} and phonon softening \cite{PRLold} in the very high 
pressure regime.

We gratefully thank S. Ciuchi for helpful discussions. M.C. acknowledges
hospitality and financial support of the Physics Department of the
University of Rome ``La Sapienza'', and of the Istituto Nazionale per la
Fisica della Materia (INFM), as well as financial support of Miur Cofin 2003.

\end{document}